# Predicting the imagined contents using brain activation

Krishna Prasad Miyapuram
Cognitive Science Program,
IIT Gandhinagar,
Ahmedabad - 382424, India
kprasad@iitgn.ac.in

Wolfram Schultz
Physiol. Devel. & Neurosci.,
University of Cambridge,
Cambridge, U.K.
ws234@cam.ac.uk

Philippe N. Tobler
Department of Economics
University of Zurich
Zurich, Switzerland
phil.tobler@econ.uzh.ch

*Abstract*— Mental imagery refers to percept-like experiences in the absence of sensory input. Brain imaging studies suggest common, modality-specific, neural correlates imagery and perception. We associated abstract visual stimuli with either visually presented or imagined monetary rewards and scrambled pictures. Brain images for a group of 12 participants were collected using functional magnetic resonance imaging. Statistical analysis showed that human midbrain regions were activated irrespective of the monetary rewards being imagined or visually present. A support vector machine trained on the midbrain activation patterns to the visually presented rewards predicted with 75% accuracy whether the participants imagined the monetary reward or the scrambled picture during imagination trials. Training samples were drawn from visually presented trials and classification accuracy was assessed for imagination trials. These results suggest the use of machine learning technique for classification of underlying cognitive states from brain imaging data.

## I. INTRODUCTION

In the sensory systems, imagery and perception of the same stimulus are known to activate common stimulus-specific brain areas. Mental imagery refers to percept-like experiences when sensory input is not present[1]. Visual imagery, for instance, is often described as "seeing with the mind's eye". Thus, when a person is creating a visual image of an apple, information which can "construct" an apple, such as its color, shape etc. is available mentally as if the person actually perceived it. Brain imaging studies suggest common, modality-specific, neural correlates imagery and perception. Mental imagery of faces activates face-perception brain areas, while imagery of places activates brain regions that are specific for place-perception [2]. The imagination and perception of tactile stimuli induces partially overlapping activations in primary and secondary somatosensory areas [3]. The imagination and presentation of odors commonly activate primary and secondary olfactory cortex [4]. The imagination of gustatory stimuli and passive gustatory perception activate the same part of the insula [5]. These findings suggest similar neural mechanisms for imagery and perception within the same modalities.

Mental images may have not only sensory but also reinforcing properties. Pelchat et al. [6] have shown that imagination of favorite foods compared to a monotonous diet activates hippocampus, insula, and caudate, regions that have been previously implicated in drug craving. On encountering a physical object, an individual's behavior of whether or not to approach this object depends on the rewarding (positive reinforcement) properties of the object. A classic paradigm to investigate reward function is Pavlovian conditioning in which an individual learns the reward-predicting properties of a previously neutral stimulus after repeated pairings of the stimulus with the reward [7]. Conceivably, Pavlovian conditioning may result in conditioned mental imagery of reward [8,9]. After learning, each presentation of the conditioned stimulus would result in a mental experience of the reward, even though the reward has not occurred yet. This conditioned reward imagery may relate to reward as the mental imagery of an object relates to its perception [8-10].

To determine whether the common activation found in imagined and visually presented trials were sufficiently strong and reliable, we used a pattern classification approach to predict whether the participants were imagining a reward or a no reward. Previous studies used simple inspection of time courses [2], correlation between patterns [11], linear discriminant analysis [12,13], and support vector machines [13-17] for predicting brain states. This approach, popularly known as mind-reading (see [18] for a review) is particularly suitable for our reward imagery study. In this study we choose a simple support vector machine (SVM) trained on the data from visual presentation trials to predict the cognitive states from the imagination trials.

## II. MATERIALS AND METHODS

### A. Experimental design

Four visual stimuli were associated with visual presented or imagined money (picture of a UK £20 note) or a scrambled picture. In each trial, a fixation '+' symbol was shown on the screen centrally for a variable inter-trial interval with a mean of 4 sec. Subsequently a conditioned stimulus (CS) appeared for 2 sec in the center of the screen. In the visual presentation trials, the CS was followed by 1 sec presentation of a £20 money bill or a scrambled picture, whereas in the imagination trials a blank screen followed the CS. Participants reported what was presented or the what they imagined within the next

This research was supported by Cambridge Nehru Fellowship (KPM), Swiss National Science Foundation (PNT) and Wellcome Trust, UK (WS).

1.5 sec by pressing a button corresponding to three response options (money bill; scrambled picture; nothing), whose spatial location was varied randomly on the screen from trial-to-trial.

*B. Data Acquisition and Analysis*

Twelve (5 female) right handed participants (age range 21.8 - 30.5, mean 25.5 years) were tested. Functional imaging was performed on a MedSpec (Bruker, Ettlingen, Germany) scanner at 3 Tesla. Standard single-subject and group analysis using statistical parametric mapping was applied to functional brain images [19-21]. A conjunction analysis was performed using the minimum statistics approach [22] to test for brain regions commonly activated preferentially to the reward-predicting stimulus in both the visual presentation and imagination trials. The results of brain imaging data are presented elsewhere [23]. This paper presents non-overlapping results using machine learning techniques.

Based on previous primate and human brain imaging studies [24-26], we defined a region of interest consisting of substantia nigra on both sides of the brain extended by 2 mm using the WFU Pick Atlas toolbox [27,28]. We extracted the parameter estimates (betas) at peak coordinates of midbrain activation from the general linear model corresponding to the four abstract stimuli. We used the midbrain betas from the visual presentation trials to train a linear support vector machine (SVM) classifier [29] to find the optimal hyper-plane discriminating the money bill and the scrambled picture. The midbrain betas from the imagination trials were used to test the classifier. We used the OSU SVM Toolbox for Matlab (http://sourceforge.net/projects/svm/) based on the LIBSVM package (http://www.csie.ntu.edu.tw/~cjlin/libsvm/) for training and testing of the SVM classifier.

The SVM is particularly suitable for high dimensional data and yields the optimal hyperplane that discriminates between reward and no reward states. The SVM is therefore better than a linear discriminant analysis, because a number of hyperplanes can be found to separate the reward from no reward, while the SVM finds the optimal one. The performance of this classifier was assessed using a Receiver Operating Characteristic (ROC) curve (see [30] 2006 for algorithm to generate ROC curve). To plot the ROC curve, we used the decision values of the classifier that suggested the strength with which each particular instance was classified as the stimulus predicting money bill (positive class) or that predicting scrambled picture (negative class). The ROC curve depicts the true positive rate (correct classification of a sample from positive class, sensitivity) of the classifier as a function of false positive rate (incorrect classification of a sample from negative class, 1- specificity). The findings from the classifier prediction analysis are expected to be a strong addition to the claim that imagination and visual presentation involve similar brain areas.

III. RESULTS

Using all the voxels in midbrain activation cluster, we found 75% correct classification for both the stimuli predicting the money bill and for the stimuli predicting the scrambled picture. The classification accuracy was significantly greater than chance level (chi square = 6.171, p<0.05). The performance of this classifier was assessed using a Receiver Operating Characteristic (ROC) curve. The chance level prediction on the ROC curve is a line with 45 degrees from the origin. The ROC curve was found to be on the upper half diagonal with 0.78 area under the curve confirming better than chance level prediction. Thus, the midbrain activation could reliably classify reward from no reward in the imagination trials based on training examples from the perception trials. Probability estimates of midbrain activation reflecting reward was obtained by fitting a sigmoid curve using the decision values of the classifier [30].

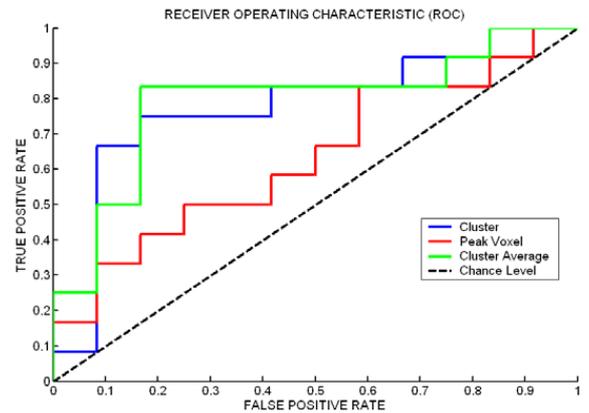

Fig. 1. Receiver operating Characteristic curve for three instances of support vector machine classifier using all voxels in cluster, peak voxel only, and the average of all voxels in cluster.

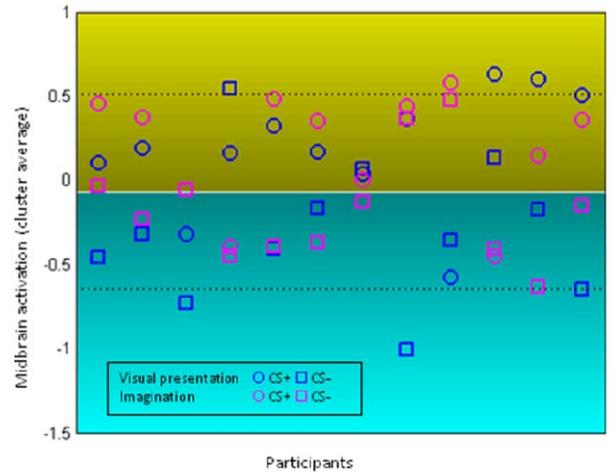

Fig. 2. Classifier output for training examples (visual presentation) and test data (imagination trials) for discriminating reward pictures (CS+) from scrambled pictures (CS-) for each participant.

IV. CONCLUSION

The results of classifier prediction strengthen the finding of common neural substrate in the midbrain for visually presented and imagined rewards. The classifier trained on midbrain activation from visually presented trials could successfully decode whether the participants were imaging reward or no reward. Further from the ROC analysis used to evaluate the performance of the classifier, it is found that using the voxel values from the entire cluster (or an average value of the cluster activation) was superior to using the peak voxel alone. Using a support vector machine as a classifier was suitable for high dimensional data as well as in obtaining the best discriminant function. While the classifier was used successfully across participants, a limitation of this analysis was that a finer analysis on a single trial basis was not performed. Another limitation of our study is that only Support vector machine was used as a classifier algorithm. Other machine learning approaches should be explored to compare across classifiers [32].